\documentstyle[12pt,epsf,epsfig,cite]{article}

\newcommand{\lesssim}{ \mathop{}_{\textstyle \sim}^{\textstyle <} }

\begin{document}
\baselineskip 0.6cm

\hfill TU-726

\hfill hep-ph/0408138

\begin{center}
{\large\bf Measurement of Squark Flavor Mixings in Supersymmetric Models 
at Super $B$ Factory}

\vspace{10mm}
{\bf Motoi Endo\footnote{E-mail: endo@tuhep.phys.tohoku.ac.jp}}
 and {\bf Satoshi Mishima\footnote{E-mail: mishima@tuhep.phys.tohoku.ac.jp}}

\vspace{5mm}
{\it Department of Physics, Tohoku University, Sendai 980-8578, Japan} 
\end{center}

\vspace{25mm}
\begin{abstract}
We discuss potential to measure squark flavor mixings based on future 
data at super $B$ factory and LHC.  In particular we focus on the imaginary 
part of the mixings by investigating the CP violating observables.  
As a result, we find they are determined with the uncertainty about 
10 \% at best.  
\end{abstract}

\newpage

The naturalness problem inherent in the standard model (SM) implies new 
physics beyond the SM at the electroweak scale.  Low scale supersymmetry 
is one of the most promising solutions to the problem.  We expect to detect 
and measure such a new physics at TeV scale by high energy collider 
experiment at LHC in the near future.  In the case of supersymmetry models, 
LHC might discover some colored supersymmetric particles and establish the 
presence of supersymmetry or its extension.  In such an era, we shall proceed 
to the next step, the determination of model parameters.  This is inevitable 
for further study of physics at high energy scale, including flavor structure 
and its origin.  

The flavor-changing-neutral-current (FCNC) is one of the most attractive 
phenomena to investigate supersymmetry and further physics at higher energy 
scale.  In fact FCNC in the SM is sufficiently suppressed by the GIM 
mechanism.  In contrast, the minimal supersymmetric standard model (MSSM) 
has the additional particles, and the superpartner of matters has 
flavor mixings in the mass matrix.  FCNC is then sensitive to these mixings 
and becomes large generically.  The flavor mixings between first two 
generations are strongly constrained from the $K^0-\overline{K}^0$ mixing, 
and $1-3$ mixings are also limited experimentally.  On the other hand, the 
supersymmetry contributions to squark flavor changing amplitudes between 
$2-3$ generations, in which we study especially the $b \to s$ transition here, 
may be large even when we consider the experimental constraint from the 
inclusive branching ratio of $b \to s \gamma$ and the neutron and atomic 
EDMs~\cite{Hisano:2003iw,Endo:2003te,Hisano:2004tf}.  
That is, the $b \to s$ transition is useful to study models of supersymmetry.  

The squark mixings generally have an ${\cal O}(1)$ phase and induce 
CP violation.  Furthermore as noted later, the processes 
which have both violations of CP and flavor are much sensitive to the 
imaginary part of the squark mixings.  Thus we expect it to be possible 
to measure those parameters without suffering from large uncertainties 
in future experiments.  

In addition, one of the most attractive processes which include both 
CP violating and flavor changing is the mixing-induced CP asymmetry for 
$B_d\to \phi K_S$.  Belle has reported large deviation, 
$S_{\phi K_S} = -0.96 \pm 0.50^{+0.09}_{-0.11}$~\cite{Abe:2003yt}, 
from the SM prediction, which is $3.5\sigma$ away, though BABAR result 
is $S_{\phi K_S} = 0.47 \pm 0.34^{+0.08}_{-0.06}$~\cite{Aubert:2004ii} 
in agreement with the SM.  Indeed the situation is not settled yet, however 
if the Belle result is correct, it is not only the first signal for 
new physics through $b \to s$ FCNC processes, but also it implies large 
CP violating phase in the scalar bottom to strange flavor mixings~\cite{phiKS,Khalil:2003bi,Hisano:2003iw,Endo:2004xt}.  
Therefore in this letter we focus on the situation that CP symmetry 
is almost maximally violated in the $b \to s$ squark mixings.  

Although the masses of the supersymmetric particles will be measured 
at LHC, in order to study the flavor mixings in the squark mass matrices 
we need some experiments with high luminosity.  Particularly super $B$ 
factory with the luminosity of $10^{35}\, {\rm cm}^{-2}{\rm sec}^{-1}$ is 
very useful for the measurement of the $b \to s$ mixings.  Actually various 
FCNC processes with CP violation are proposed to be measured 
with high accuracy.  In this letter we focus on the CP violating processes, 
the mixing-induced CP asymmetry of $B_d\to \phi K_S$ ($S_{\phi K_S}$), of 
$B_d\to \eta' K_S$ ($S_{\eta' K_S}$), of $B_d\to K^* \gamma$ ($S_{K^*\gamma}$), 
the direct CP asymmetry of $b\to s \gamma$ ($A_{\rm CP}^{b\to s\gamma}$) 
and the $B_s-\overline{B_s}$ mixing ($\Delta M_s$), after one year of 
running at super $B$ factory.  

So far many studies for the supersymmetric contributions to the 
$b \to s$ transitions have been made.  Some are interested in potential 
to detect the MSSM and some flavor models through the transition processes.  
The authors in Ref.~\cite{Goto:2003iu} investigated the processes in systematic 
way for some typical flavor models in supersymmetry.  Their study is not only 
to search potential for detection but also to identify these models at 
super $B$ factory and they showed that the processes are very helpful.  
However this argument is based on model dependent analysis.  
Rather for the purpose of investigation of higher scale physics more 
definitely, we need model independent one.  
Therefore in this letter we do not assume any flavor models, and study the 
$b \to s$ transition processes in model independent way.  

The mass insertion approximation (MIA)~\cite{Hall:1985dx,Gabbiani:1996hi} 
is a powerful technique for model independent analysis of the MSSM.  
When off-diagonal elements in the squark mass matrices are small enough, 
the squark propagators with $\tilde b\to \tilde s$ FCNC can be expanded 
as a series in terms of 
\begin{eqnarray}
(\delta^d_{LL})_{23} = \frac{(m^2_{\tilde d_{L}})_{23}}{m^2_{\tilde q}}
\;,&~~~&
(\delta^d_{RR})_{23} = \frac{(m^2_{\tilde d_{R}})_{23}}{m^2_{\tilde q}}
\;, \nonumber\\
(\delta^d_{LR})_{23} = \frac{(m^2_{\tilde d_{LR}})_{23}}{m^2_{\tilde q}}
\;,&~~~&
(\delta^d_{RL})_{23} = (\delta^d_{LR})_{32}^*
\;,
\end{eqnarray}
where $m^2_{\tilde d}$ is the squared down-type-squark mass matrix,
$m_{\tilde q}$ an averaged squark mass.  
There is chirality structure in the squark mass matrices.  
It is important that these parameters generically 
have an ${\cal O}(1)$ phase and imaginary part of them induces CP and 
flavor violations simultaneously.  
In this letter we discuss potential to constrain and measure these MIA 
parameters by the data of the $b \to s$ transition processes which will 
be measured in future at super $B$ factory and LHC.  


Here we summarize the effective Hamiltonian for $\Delta B=1$ and 
$\Delta S=1$ processes~\cite{Buchalla:1995vs}.  This is defined as 
\begin{eqnarray}
H_{\rm eff}\,
=\,
\frac{4G_F}{\sqrt{2}}
\left[
\sum_{q'=u,c}V_{q'b}V_{q's}^* 
\sum_{i=1,2}
C_i O_i^{(q')}
- V_{tb}V_{ts}^*
\sum_{i=3\sim6, 7\gamma, 8G} 
\left( C_iO_i + \widetilde{C}_i\widetilde{O}_i \right)
\right]\;,
\end{eqnarray}
where the local operators are given by
\begin{eqnarray}
& &O_1^{(q')}
 = ({\bar s}_i \gamma_\mu P_L q'_j)({\bar q'}_j \gamma^\mu P_L b_i)\;,
\;\;\;\;\;\;\;\;
O_2^{(q')}
 = ({\bar s}_i \gamma_\mu P_L q'_i)({\bar q'}_j \gamma^\mu P_L b_j)\;,
\nonumber \\
& &O_3
 =({\bar s}_i \gamma_\mu P_L b_i)\sum_{q}({\bar q}_j \gamma^\mu P_L q_j)\;,
\;\;\;\;\;\;\;
O_4
 =({\bar s}_i \gamma_\mu P_L b_j)\sum_{q}({\bar q}_j \gamma^\mu P_L q_i)\;, 
\nonumber \\
& &O_5
 =({\bar s}_i \gamma_\mu P_L b_i)\sum_{q}({\bar q}_j \gamma^\mu P_R q_j)\;,
\;\;\;\;\;\;\;
O_6
 =({\bar s}_i \gamma_\mu P_L b_j)\sum_{q}({\bar q}_j \gamma^\mu P_R q_i)\;, 
\nonumber \\
& &O_{7\gamma} 
 =  \frac{e}{16\pi^2} m_b {\bar s}_i \sigma^{\mu\nu}
   P_R b_i F_{\mu\nu}\;,
\;\;\;\;\;\;\;\;\;
O_{8G}    
 =  \frac{g_s}{16\pi^2} m_b {\bar s}_i \sigma^{\mu\nu}
   P_R T^a_{ij} b_j G^a_{\mu\nu}\;,
\label{eq:localop}
\end{eqnarray} 
with $P_R = (1+\gamma_5)/2$ and $P_L = (1-\gamma_5)/2$.
Here, $i$ and $j$ are color indices, and $q$ is taken to be $u$, $d$, $s$
and $c$.  The terms with tilde are obtained by flipping chiralities, 
$L \leftrightarrow R$, in Eq.~(\ref{eq:localop}). 
The gluino contributions to the Wilson coefficients $C_i$ 
at supersymmetry scale $M_S$ are calculated as 
\begin{eqnarray}
  C_3^{\tilde g} (M_S)
  &\simeq&
  \frac{\sqrt{2} \alpha_s^2}{4G_F V_{tb} V_{ts}^*  m_{\tilde{q}}^2}  
  (\delta_{LL}^d)_{23} \left[ -\frac{1}{9} B_1(x) - \frac{5}{9} B_2(x)
    - \frac{1}{18} P_1(x) -\frac{1}{2} P_2(x) \right]
  \;,\nonumber\\
  C_4^{\tilde g}(M_S)
  &\simeq&
  \frac{\sqrt{2} \alpha_s^2}{4G_F V_{tb} V_{ts}^*  m_{\tilde{q}}^2} 
  (\delta_{LL}^d)_{23}
  \left[ -\frac{7}{3} B_1(x) + \frac{1}{3} B_2(x) + \frac{1}{6} P_1(x)
    +\frac{3}{2} P_2(x) \right]
  \;,\nonumber\\
  C_5^{\tilde g} (M_S)
  &\simeq&
  \frac{\sqrt{2} \alpha_s^2}{4G_F V_{tb} V_{ts}^* m_{\tilde{q}}^2} 
  (\delta_{LL}^d)_{23}
  \left[ \frac{10}{9} B_1(x) + \frac{1}{18} B_2(x) - \frac{1}{18} P_1(x)
    -\frac{1}{2} P_2(x) \right]
  \;,\nonumber\\
  C_6^{\tilde g} (M_S)
  &\simeq&
  \frac{\sqrt{2} \alpha_s^2}{4G_F V_{tb} V_{ts}^*  m_{\tilde q}^2} 
  (\delta_{LL}^d)_{23}
  \left[ -\frac{2}{3} B_1(x) + \frac{7}{6} B_2(x) + \frac{1}{6} P_1(x)
    +\frac{3}{2} P_2(x) \right]
  \;,\nonumber\\
  C_{7\gamma}^{\tilde g} (M_S)
  &\simeq&
  -\frac{\sqrt{2} \alpha_s \pi}
  {6G_F V_{tb} V_{ts}^*  m_{\tilde q}^2}
  \left[
    (\delta_{LL}^d)_{23}\, 
    \left(
      \frac{8}{3} M_3(x)
      - 
      \mu_H \tan\beta \frac{m_{\tilde g}}{m_{\tilde q}^2}
      \frac{8}{3} M_a(x)
    \right)
  \right.\nonumber\\
  & & \left.\ \ \ \ \ \ \ \ \ \ \ \ \ \ \ \ \ \ \ 
    +
    (\delta_{LR}^d)_{23}\, \frac{m_{\tilde g}}{m_b}\,
    \frac{8}{3} M_1(x) \right]
  \;,\nonumber\\
  C_{8G}^{\tilde g} (M_S)
  &\simeq&
  -\frac{\sqrt{2} \alpha_s \pi}
  {2G_F V_{tb} V_{ts}^*  m_{\tilde q}^2}
  \left[
    (\delta_{LL}^d)_{23}
    \left\{
      \left( \frac{1}{3} M_3(x) + 3 M_4(x)\right)
    \right.\right.\nonumber\\
  & &\left.\left. \hspace{-5mm}
      - 
      \mu_H \tan\beta \frac{m_{\tilde g}}{m_{\tilde q}^2}
      \left(\frac{1}{3} M_a(x)+ 3M_b(x)\right)
    \right\}
    +\,
    (\delta_{LR}^d)_{23}\, \frac{m_{\tilde{g}}}{m_b}
    \left(\frac{1}{3} M_1(x) + 3 M_2(x)\right)\right]\;.
\end{eqnarray}
Here $B_{1,2}$, $P_{1,2}$, $M_{1-4}$ and $M_{a,b}$ are the loop
functions\footnote{The loop functions $B_{1,2}$, $P_{1,2}$ and
$M_{1-4}$ are defined in Ref.~\cite{Gabbiani:1996hi}. $M_{a}$ and
$M_{b}$ are the same as $M_{1}$ and $M_{2}$ in Ref.~\cite{Hisano:2004tf},
respectively.} and $x = m^2_{\tilde{g}}/m^2_{\tilde{q}}$, where 
$m_{\tilde g}$ is the gluino mass.  The superscript $\tilde g$ denotes 
the coefficients come from the gluino contributions.  The other 
contributions include those from the SM, the charged Higgs, the chargino 
and the neutralino.  We note that these contributions except for gluino 
one have no extra CP violating phases once we consider the experimental 
constraint from the electron, neutron and atomic EDMs.  Consequently, 
the CP violating and flavor changing phenomena are generally less 
sensitive to these operators and controlled by the gluino contributions.  

The penguin coefficients $C_{3-6}^{\tilde g}$ depend only 
on $(\delta^d_{LL})_{23}$, on the other hand the dipole-penguin coefficients 
$C_{7\gamma}^{\tilde g}$ and $C_{8G}^{\tilde g}$ on both $(\delta^d_{LL})_{23}$ 
and $(\delta^d_{LR})_{23}$.  
In particular that of the $(\delta^d_{LL})_{23}$ contributions 
to $C_{7\gamma}^{\tilde g}$ and $C_{8G}^{\tilde g}$ are enhanced by a ratio 
of the Higgs VEVs, $\tan\beta$, 
because of the double-mass-insertion diagrams.  This type of diagrams is 
composed of $(\delta^d_{LL})_{23}$ and $(\delta^d_{LR})_{33}$.  
We note the diagram is very similar to one of $(\delta^d_{LR})_{23}$.  
To see this argument more clearly, let us denote the Wilson coefficients as 
\begin{eqnarray}
  C_{7\gamma}^{\tilde g} \propto
  \left[(\delta^d_{LL})_{23} + c^{7\gamma} (\delta^d_{LR})_{23}\right],
  ~~~
  C_{8G}^{\tilde g} \propto 
  \left[(\delta^d_{LL})_{23} + c^{8G} (\delta^d_{LR})_{23}\right],
\end{eqnarray}
and define ``total mixing'', 
\begin{eqnarray}
  (\delta^d_{LL})_{23}^{\rm (tot)} \equiv (\delta^d_{LL})_{23} 
  + c^{7\gamma, 8G} (\delta^d_{LR})_{23}.
  \label{eq:totalmixing}
\end{eqnarray}
Here the coefficient $c$'s are estimated as 
$c \simeq (\delta^d_{LR})_{33}^{-1}$ up to the loop functions.  
It is important that $c$'s satisfy $c^{8G}/c^{7\gamma} 
( \simeq \tilde c^{8G}/\tilde c^{7\gamma} ) \simeq 1$ 
within an error about 40 \%.  If this relation is exact, 
the result from the contribution with $(\delta_{LL}^d)_{23}$ become the 
same as that with $(\delta_{LR}^d)_{23}$ in the processes dominated by 
$C_{7\gamma}^{\tilde g}$ and $C_{8G}^{\tilde g}$.  
That is, we can study $(\delta_{LL}^d)_{23}$ and $(\delta_{LR}^d)_{23}$ 
simultaneously, and $(\delta_{RR}^d)_{23}$ and $(\delta_{RL}^d)_{23}$, too.  
In this letter we first investigate the LL 
and RR squark mixings and then discuss to distinguish LR and RL from them.  


Let us review the processes which we consider in this letter.  
The $b \to s$ effective Hamiltonian induces a lot of FCNC processes.  
The important observables in this letter are the following CP violating ones: 
$S_{\phi K_S}$, $S_{\eta' K_S}$, $S_{K^*\gamma}$ and 
$A_{\rm CP}^{b\to s\gamma}$.  
The supersymmetry contributions can be comparable to or even 
larger than that of the SM, and depend dominantly on the dipole moment 
operators, $C_{7\gamma}$ and $C_{8G}$.  We determine the the imaginary 
part of the ``total'' squark mixings by these processes.  

The $B_s-\overline{B_s}$ mixing, $\Delta M_s$, is also interesting here.  
$\Delta M_s$ depends on the four-quark operators and the double-mass-insertion 
diagrams do not dominate.  
As a result, it is possible to distinguish the LL and RR mixings 
from the others by this process.  As will be mentioned later 
only a product of $(\delta^d_{LL})_{23}$ and $(\delta^d_{RR})_{23}$ induces 
large $\Delta M_s$~\cite{Ball:2003se}, 
which may be detected at LHC~\cite{Ball:2000ba,LHCb:tech}.  

Too large MIA parameters however exceed the current bounds from 
${\rm Br}(b\to s \gamma)$ and the contribution of the strange quark color 
electric-dipole-moment operator (CEDM) to the neutron and atomic EDMs.  
We consider these processes as a constraint on the parameter space.  

We comment on the model parameters in this letter.  Here the $b \to s$ 
transition processes are selected so that a number of the parameters is 
as small as possible.  In fact, the relevant soft parameters are the squark 
masses $m_{\tilde q}$, the gluino mass $m_{\tilde g}$, the higgsino mass 
parameter $\mu_H$ and $\tan\beta$ (and top trilinear coupling for the 
contribution to the strange CEDM from chargino mediated 
diagram~\cite{Endo:2003te}).  Also the real part of the Wilson coefficients 
includes the neutralino and chargino masses and the charged Higgs mass.  

Before proceeding to each modes, we comment on the other FCNC processes.  
There are attractive observables other than those in this letter.  
For example, $B_s \to l\,l$ is triggered by mediating the Higgses 
and may become sizable~\cite{Babu:1999hn}.  
However the supersymmetry contributions to 
the observables depend strongly on additional model parameters like 
Higgs mass ones.  Thus we do not consider those processes in this letter.  

The branching ratio of the inclusive decay $b \to s \gamma$ is one of 
the most important processes when we consider the $b \to s$ transition.  
In fact the branching ratio is proportional to the sum of the squared Wilson 
coefficients, $(|C_{7\gamma}(m_b)|^2+|\widetilde{C}_{7\gamma}(m_b)|^2)$, 
at the leading order.  
And the SM prediction of the branching ratio, which can be calculated 
cleanly, is now in agreement with experimental data~\cite{Barate:1998vz} 
within errors.  In consequence, too large $b \to s$ transition amplitude 
is excluded by considering ${\rm Br}(b \to s \gamma)$~\cite{Kagan:1998ym}.  
The theoretical prediction has been calculated at next-to-leading order in 
renormalization group improved perturbation theory and its error is
about 10 \%~\cite{Hurth:2003dk}. It is expected to be reduced down to
5 \% within a few years when all next-to-next-to-leading-order
corrections are included~\cite{Hurth:2003dk}. 
The experimental error will be also reduced 10 \% to 
5 \% at super $B$ factory after one year of running~\cite{Akeroyd:2004mj}.
Despite of these expectations, we consider ${\rm Br}(b \to s \gamma)$ 
just as a constraint on the parameter space and are less interested in 
detection ability of supersymmetry in this letter.  Thus we take a rather 
conservative value, 
\begin{eqnarray}
  2.0 \times 10^{-4} < {\rm Br}(b \to s \gamma) < 4.5 \times 10^{-4},
\end{eqnarray}
instead of improved value which we expect in the near future.  

The MIA parameters are also constrained by the EDMs, though there are 
large hadronic uncertainties left.  The squark mixings contribute to the 
strange quark CEDM and the CEDM is limited by the neutron and atomic EDMs.  
The neutron EDM now gives the strongest bound~\cite{Hisano:2004tf},
\begin{eqnarray}
  e|\tilde d_s| < 1.9(2.4) \times 10^{-25}\ e{\rm cm},
\end{eqnarray}
where the number in the parentheses represents the assumption of the 
Pecci-Quinn (PQ) symmetry.  The measurement of the deuteron EDM is also 
proposed and its sensitivity is $\sim 10^{-26}$ $e$cm.  
As a result, at first we find that the gluino 
diagram contributions generically impose strong correlations on the 
squark mixings~\cite{Hisano:2003iw,Hisano:2004tf}, 
\begin{eqnarray}
  \sqrt{|{\rm Im}(\delta^d_{LL})_{23}(\delta^d_{RR})_{32}|} &\lesssim& 
  1.6 (2.0) \times 10^{-4}, \nonumber\\
  \sqrt{|{\rm Im}(\delta^d_{LL})_{23}(\delta^d_{LR})_{32}|}
  ~~~{\rm and}~~~ \sqrt{|{\rm Im}(\delta^d_{LR})_{23}(\delta^d_{RR})_{32}|} 
  &\lesssim& 4.4 (5.6) \times 10^{-6},
\end{eqnarray}
for the soft masses $m_{\tilde g} = m_{\tilde q} = 500$ GeV and 
$\mu_H \tan\beta = 5000$ GeV.  In this letter we thus assume that 
the phases of the squark mixings are perfectly aligned.  
Finally the chargino diagram also gives a sizable contribution to the 
strange CEDM~\cite{Endo:2003te}.  The contribution leads to a constraint 
on the LL squark mixings, $|{\rm Im}(\delta^d_{LL})_{23}| \lesssim 10^{-1}$, 
in the generic situation.  It is comparable to ${\rm Br}(b \to s \gamma)$, 
and detailed analysis shows that it is rather parameter 
dependent~\cite{Endo:2004xt}.

$B_d\to \phi K_S$ and $B_d\to \eta' K_S$ are very interesting decay modes 
in the search for new physics~\cite{phiKS,Khalil:2003bi,Hisano:2003iw,Endo:2004xt}. 
In the SM the both mixing-induced CP asymmetries for $\phi K_S$ and $\eta'
K_S$ are equal to $\sin(2\phi_1)$, which is measured by a tree dominated 
process $B_d\to J/\Psi K_S$. Any deviation larger than ${\cal O}
(\lambda^2)$ would be a signal for new physics~\cite{Grossman:1996ke}.  
The mixing-induced CP asymmetry for a CP eigenstate $f_{\rm CP}$ is given by
\begin{eqnarray}
S_{f_{\rm CP}}\,
=\,
\xi_{f_{\rm CP}}
\frac
{2\; {\rm Im}\left[
e^{-2i\; \phi_1} A(\overline{B_d}\to f_{\rm CP})/A(B_d\to f_{\rm CP})
 \right]}
{|A(\overline{B_d}\to f_{\rm CP})/A(B_d\to f_{\rm CP})|^2+1}
\;,
\end{eqnarray}
with ${\rm CP}|f_{\rm CP}\rangle = \xi_{f_{\rm CP}}|f_{\rm CP}\rangle$. 
New CP violating phases drive the mixing-induced CP asymmetry away from 
the SM prediction.  In the supersymmetry models the deviation is induced by 
the additional phases of the squark mixings.  The dominant supersymmetric 
contribution comes from the chromo-magnetic penguin, which depends on the 
squark mixings, and thus induce the deviation of the mixing-induced CP 
asymmetry.  
The supersymmetric contribution to the amplitudes in $\phi K_S$ is denoted as,
\begin{eqnarray}
A^{\rm SUSY}(\overline{B_d}\to \phi K_S)
\,
\propto\,
C_{8G}^{\tilde g}(m_b)\, +\, \widetilde{C}_{8G}^{\tilde g}(m_b)
\;.
\end{eqnarray}
In the case of $\eta' K_S$ it comes from the chromo-magnetic penguin 
as that of $\phi K_S$, however the chirality structure is different:
\begin{eqnarray}
A^{\rm SUSY}(\overline{B_d}\to \eta' K_S)
\,
\propto\,
C_{8G}^{\tilde g}(m_b)\, -\, \widetilde{C}_{8G}^{\tilde g}(m_b)
\;,
\end{eqnarray}
because of parity difference between $\phi$ and $\eta'$~\cite{Khalil:2003bi}. 
Consequently, we can classify the chromo-magnetic penguin operators, 
in particular whether $\widetilde C_{8G}$ is large or not by using 
the both data of $S_{\phi K_S}$ and $S_{\eta' K_S}$.
The current results are $S_{\eta' K_S} =
0.02 \pm 0.34 \pm 0.03$  at BABAR~\cite{Aubert:2003bq} and $0.43 \pm
0.27 \pm 0.05$ at Belle~\cite{Abe:2003yt}.
Although $S_{\eta' K_S}$ is slightly smaller than the SM prospect, the
deviation from $S_{\phi K_S}$ is not yet measured.
The experimental errors in $S_{\phi K_S}$ and $S_{\eta' K_S}$ will be
less than $0.1$ at super $B$ factory so that the
discrepancy may be observed in future.

Although the experimental signal is clean, theoretical estimation includes 
large uncertainties.  They originate in the calculations of hadronic matrix 
elements.  To obtain the matrix elements, there are several
approaches, naive factorization~\cite{Wirbel:1985ji}, generalized
factorization~\cite{Ali:1997nh}, QCD factorization~\cite{Beneke:1999br},
perturbative QCD~\cite{Keum:2000ph}, and so on.  Each method is however plagued 
with large theoretical uncertainties and it is expected that the estimation 
will be improved and they will be reduced in future.
As a reference, let us use the generalized factorization method.  
Then the main theoretical error comes from the matrix element of 
the chromo-magnetic penguin, 
\begin{eqnarray}
\langle \phi K_S | O_{8G} | \overline{B_d} \rangle\,
=\,
- 
\frac{\alpha_s(m_b)}{4\pi}\,
\frac{m_b}{\sqrt{q^2}}
\left\langle \phi K_S \left|
O_4 + O_6 -\frac{1}{3}(O_3 + O_5)
\right| \overline{B_d} \right\rangle
\;,
\end{eqnarray}
where $q^2$ is the momentum transferred by the gluon in $O_{8G}$.  Though the 
parameter $q^2$ is ambiguous in the generalized factorization, this is not an 
inherent uncertainty in the estimation of $S_{\phi K_S}$.  In fact this can be 
removed once we apply QCD factorization~\cite{Beneke:1999br} or perturbative 
QCD~\cite{Mishima:2003wm}.  Rather the method in this letter has an 
advantageous of simplicity in the analysis.  
Here we take $q^2=(M_B^2-M_\phi^2/2+M_K^2)/2$~\cite{Ali:1997nh}.  
This argument is also applied for the analysis of $\eta' K_S$ by replacing 
$\phi$ with $\eta'$.

$S_{K^* \gamma}$ is one of the important observables for the measurement  
of the squark mixings in $\widetilde C_{7\gamma}$. 
We measure the asymmetry of $B_d$ decay into a CP eigenstate
$K^{*0}\gamma (\to K_S\pi^0\gamma)$~\cite{Atwood:1997zr}.  
The leading contribution in $B_d\to K^{*}\gamma$ is the photo-magnetic penguin.
Other contributions, which come from $O_2$ and $O_{8G}$, are 
sub-leading~\cite{Bosch:2001gv,Keum:2004is} and neglected
in the following study for simplicity.  
In consequence, $S_{K^* \gamma}$ is generated from the interference 
between $C_{7\gamma}$ and $\widetilde{C}_{7\gamma}$ terms: 
\begin{eqnarray}
S_{K^* \gamma}\, 
=\, 
\frac{2\; {\rm Im}\left[
e^{-2i\; \phi_1}\widetilde{C}_{7\gamma}(m_b)/C_{7\gamma}(m_b) \right]}
{\left|\widetilde{C}_{7\gamma}(m_b)/C_{7\gamma}(m_b)\right|^2+1}
\;.
\end{eqnarray}
$S_{K^* \gamma}$ vanishes in the SM since it is suppressed by 
$2m_s/m_b$~\cite{Atwood:1997zr}.  
However if there is a sizable parameter insertion in $\widetilde C_{7\gamma}$, 
the deviation of $S_{K^* \gamma}$ can be large in the MSSM.  At the leading 
level theoretical uncertainty cancels out in the ratio, because 
the matrix elements for $O_{7\gamma}$ is the same as that for 
$\widetilde{O}_{7\gamma}$. This is an advantage of $S_{K^* \gamma}$ 
compared with $S_{\phi K_S}$ and $S_{\eta' K_S}$.  The current data is 
$S_{K^*\gamma} = 0.25 \pm 0.63 \pm 0.14$ at BABAR~\cite{Aubert:2004pe}, 
which is still ambiguous. The error will be
reduced to about $0.1$ at super $B$ factory~\cite{Akeroyd:2004mj}.

The direct CP asymmetry for $b\to s\gamma$ is sensitive to the LL and LR 
mixings.  This is estimated at the leading order as
\begin{eqnarray}
A_{\rm CP}^{b\to s\gamma}
&=& 
\frac{1}{|C_{7\gamma}(m_b)|^2+|\widetilde{C}_{7\gamma}(m_b)|^2}\,
\Bigg[
    a_{27}\,{\rm Im}\left(C_2(m_b) C_{7\gamma}^*(m_b)
    +\widetilde{C}_2(m_b) \widetilde{C}_{7\gamma}^{*}(m_b)\right)
\nonumber\\
& & \hspace{35mm}
    +\, a_{28}\,{\rm Im}\left(C_2(m_b) C_{8G}^*(m_b)
    +\widetilde{C}_2(m_b) \widetilde{C}_{8G}^{*}(m_b)\right)
\nonumber\\
& & \hspace{35mm}
    +\,  a_{87}\,{\rm Im}\left(C_{8G}(m_b) C_{7\gamma}^*(m_b)
    +\widetilde{C}_{8G}(m_b) \widetilde{C}_{7\gamma}^{*}(m_b)\right)
\Bigg]
\;,
\end{eqnarray}
with $a_{27}\sim {\cal O}(10^{-2})$, $a_{28}\sim {\cal O}(10^{-3})$ 
and $a_{87} \sim {\cal O}(10^{-1})$~\cite{Kagan:1998bh}.  Although the 
analysis at the next-to-leading order is given in Ref.~\cite{Hurth:2003dk}, 
we use the result at the leading level for simplicity.  
The direct CP asymmetry requires at least two
amplitudes with different weak phases and different strong phases.
In the SM, there is no relative weak phase between $C_{7\gamma}$ and $C_{8G}$ 
while $C_2$ has a different one.  
Consequently, the SM prediction due to the interference between $C_2$ and 
$C_{7\gamma}$ is estimated as ${\cal O}(10^{-1})$ \%~\cite{Hurth:2003dk}.  
On the other hand, $\widetilde C_2$, $\widetilde C_{7\gamma}$ and 
$\widetilde C_{8G}$ are suppressed sufficiently in the SM.  
The current experimental data, $0.025\pm 0.050\pm 0.015$ 
at BABAR~\cite{Aubert:2004hq} and $0.002\pm 0.050\pm 0.030$ 
at Belle~\cite{Nishida:2003yw}, are consistent with null asymmetry.  
Their errors will be reduced to less than 1 \% 
at super $B$ factory~\cite{Akeroyd:2004mj}.  
If there is a sizable contribution from the squark mixings on $C_{7\gamma}$ 
and $C_{8G}$, the interference between them can generate larger CP asymmetry, 
at most 10 \%, considering that $a_{87}$ is much larger than the others. 
However the supersymmetry contributions to $\widetilde C_2$, $\widetilde 
C_{7\gamma}$ and $\widetilde C_{8G}$ except for that from the gluino diagrams 
are again negligible.  As a result, though the gluino contribution may be 
sizable, the phase of $\widetilde C_{7\gamma}$ aligns with that of 
$\widetilde C_{8G}$ and the CP asymmetry induced by $\widetilde C$ sector 
is still small.  

The $B_s-\overline{B_s}$ mixing is helpful to distinguish the mass 
insertions scheme, that is, a product of the LL and RR mixings, LL+RR, 
and the others.  The gluino contributions to the mass difference between 
the mass eigenstates of the $B_s -\overline{B_s}$ system is given by
\begin{eqnarray}
\frac{\Delta M_s^{\tilde{g}}}{\Delta M_s^{\mathrm{SM}}} 
&=& 
a_1
\left[(\delta^d_{LL})_{23}^2 + (\delta^d_{RR})_{23}^2 \right] 
+ a_2 
\left[(\delta^d_{LR})_{23}^2 + (\delta^d_{RL})_{23}^2 \right] 
\nonumber\\
&& 
+ a_3 \left[(\delta^d_{LR})_{23} (\delta^d_{RL})_{23} \right] 
+ a_4 \left[(\delta^d_{LL})_{23} (\delta^d_{RR})_{23} \right]
\;,
\end{eqnarray}
with $a_1\sim 1.44$, $a_2\sim 27.57$, $a_3\sim -44.76$ and $a_4 \sim
-175.79$ for the soft masses $m_{\tilde g} = m_{\tilde q} = 500$
GeV~\cite{Ball:2003se}.  The sizable LL+RR contribution enhances
$\Delta M_s$ significantly but the others do not.
So far there has been only the lower bound,
$\Delta M_s > 14.5\, {\rm ps}^{-1}$~\cite{exp:deltaMs}.  
On the other hand, the SM prediction 
is typically around $18\,{\rm ps}^{-1}$~\cite{Battaglia:2003in} 
where its theoretical uncertainty
coming from hadronic parameters is about 10 \% in the lattice
calculations~\cite{Yamada:2001xp}. 
In the near future, $\Delta M_s$ will be measured at Tevatron and/or LHC.  
LHC covers $\Delta M_s$ range up to $48\, {\rm ps}^{-1}$ and 95 \%
exclusion region up to $58\, {\rm ps}^{-1}$ after one
year of running~\cite{Ball:2000ba}. And expected statistical uncertainty on
$\Delta M_s$ is $0.018\, {\rm ps}^{-1}$ at $\Delta M_s=50\, {\rm
ps}^{-1}$~\cite{LHCb:tech}.


By considering the observables, $S_{\phi K_S}$, $S_{\eta' K_S}$, 
$S_{K^*\gamma}$, $A_{\rm CP}^{b\to s\gamma}$ and $\Delta M_s$, 
we study potential to constrain 
and measure the imaginary part of the squark mixings.  
Here let us summarize the setup in this analysis.  
At first we consider the situation that a large anomaly in $S_{\phi K_S}$ 
will be measured at super $B$ factory.  Actually such a large deviation is 
currently implied by the Belle result.  In order to realize large deviation 
in $S_{\phi K_S}$ the MIA parameters are favored to have the maximal phase by 
considering the constraint from ${\rm Br}(b\to s \gamma)$ and the hadronic 
EDMs.  Therefore we assume the squark mixings to be pure imaginary in the 
following analysis.  

Secondly we study the measurement ability for the squark mixings at super 
$B$ factory.  This means that we do not consider the 
uncertainties due to the calculation method of the $b \to s$ processes.  
For example, although we use the generalized factorization method for 
estimating $S_{\phi K_S}$, the technique is expected to be improved in future 
and the uncertainties may be reduced.  
In addition, other than the squark mixings we have other relevant model 
parameters.  Those parameters are assumed to be measured at LHC and we do 
not include their ambiguities in the result.  We consider only experimental 
errors as the resultant uncertainties.  

Third, some processes in this letter are planned to be measured at LHC and 
super $B$ factory.  We consider an era after one year of running at super $B$ 
factory.  Currently the proposed luminosity is not settled yet and the expected 
experimental errors for each processes at super KEKB are summarized in 
Table.~\ref{tab:superKEKB}~\cite{Akeroyd:2004mj}.  
With these values the squark mixings are determined and we show the result 
for the both cases of the luminosities.  

Fourth, the experimental uncertainty of $\Delta M_s$ is proposed to be 
negligibly small at LHC.  As a result, the theoretical uncertainty becomes 
much larger than the experimental error, unlike the other observables 
in this letter.  Thus we consider the theoretical ambiguity for $\Delta M_s$.  
We use a value 10 \% for the uncertainty in the following.  

\begin{figure}[ht]
  \begin{center}
    \includegraphics{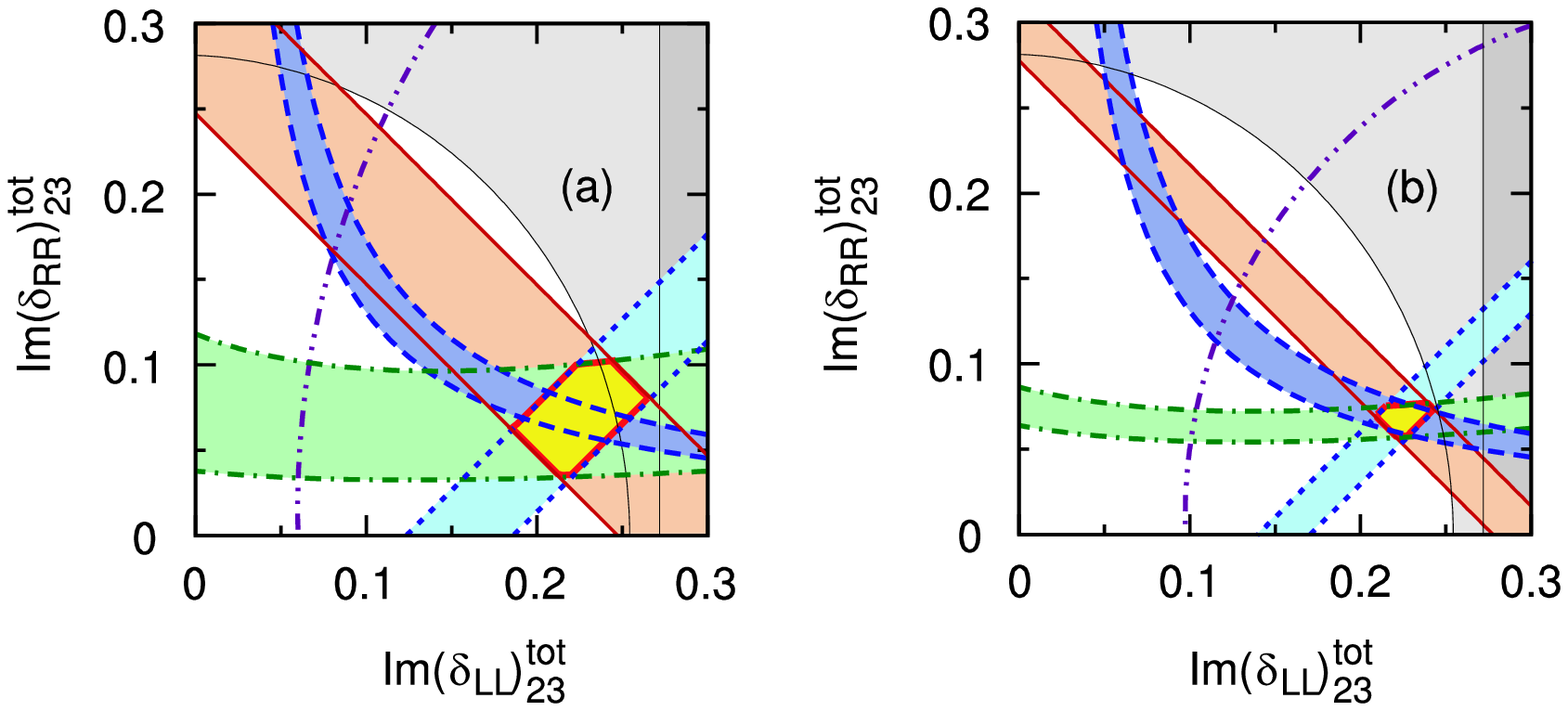}
  \end{center}
  \caption{The constant contours of $S_{\phi K_S}$(red or thick solid), 
    $S_{\eta' K_S}$(light blue or dotted), $S_{K^*\gamma}$(green or 
    dash-dotted),  $A_{\rm CP}^{b\to s\gamma}$(purple or dash double-dotted) 
    and $\Delta M_s$(blue or dashed) for changing the LL and RR squark 
    mixings.  Here we take all soft parameters $m_{\rm soft} = 500$ GeV and 
    $\tan\beta = 10$.  The bands of the experimental errors are displayed 
    at the luminosity (a) 5 ab$^{-1}$ and (b) 50 ab$^{-1}$, except for 
    $A_{\rm CP}^{b\to s\gamma}$.  Although $A_{\rm CP}^{b\to s\gamma}$ will 
    be measured at the region of larger 
    ${\rm Im}(\delta^d_{LL})_{23}^{\rm (tot)}$, 
    it is ambiguous in this parameter set.  
    We also note that $\Delta M_s$ is estimated in the condition that 
    the LL+RR squark mixings are dominant.  
    There is a region where the uncertainty band of $S_{\phi K_S}$, 
    $S_{\eta' K_S}$ and $S_{K^*\gamma}$ overlaps with each other, 
    which is filled by yellow and enclosed by red solid lines in the figures.  
    Some parameter regions are excluded experimentally: 
    the region of larger $(\delta^d_{LL})_{23}^{\rm (tot)}$ and/or 
    $(\delta^d_{RR})_{23}^{\rm (tot)}$, which is outer side of the 
    thin solid lines, 
    comes from ${\rm Br}(b \to s \gamma)$(light shadowed) and larger 
    $(\delta^d_{LL})_{23}^{\rm (tot)}$ from the EDMs(dark shadowed).}
  \label{fig:LLRR}
\end{figure}

Let us show the numerical result based on the setup given above.  
At first we assume the LR and RL squark mixings to be zero.  
Therefore the total mixings satisfy with 
$(\delta^d_{LL,RR})_{23}^{\rm (tot)} = (\delta^d_{LL,RR})_{23}$.  
Figure~\ref{fig:LLRR} displays the numerical estimation of $S_{\phi K_S}$, 
$S_{\eta' K_S}$, $S_{K^*\gamma}$ and $A_{\rm CP}^{b\to s\gamma}$.  
We show the constant contours of these observables for changing the squark 
mixings, 
$(\delta^d_{LL})_{23}$ and $(\delta^d_{RR})_{23}$.  Here we take the other 
model parameters as, the relevant soft masses (including $\mu_H$) 
$m_{\rm soft} = 500$ GeV and $\tan\beta = 10$.  Some parameter regions 
are excluded experimentally.  The region of larger $(\delta^d_{LL})_{23}$ 
and/or $(\delta^d_{RR})_{23}$ exceeds the bound from 
${\rm Br}(b \to s \gamma)$.  
Also too large $(\delta^d_{LL})_{23}$ suffers from the EDMs.  Here we show 
the bound from the neutron EDM with the PQ symmetry, though this constraint 
is weaker for this parameter set.  

In Fig.~\ref{fig:LLRR} we take a center value of each mode as 
$S_{\phi K_S} = 0.3$, $S_{\eta' K_S} = 0.5$, $S_{K^*\gamma} = -0.3$ and
$A_{\rm CP}^{b\to s\gamma} = 0.02$.  Although these are just reference, 
we choose $S_{\phi K_S}$ to have a large deviation from the SM considering 
the current result from Belle.  
Each experimental uncertainty is denoted as the band at the luminosity 
of 5 ab$^{-1}$ for Fig.~\ref{fig:LLRR} (a) and 50 ab$^{-1}$ for (b), 
except for $A_{\rm CP}^{b\to s\gamma}$.  Indeed a signal of 
$A_{\rm CP}^{b\to s\gamma}$ may be measured at super $B$ factory 
as long as ${\rm Im}(\delta^d_{LL})_{23}$ is large, the parameters here 
is not suitable to determine the squark mixings from 
$A_{\rm CP}^{b\to s\gamma}$, because of the wide extent of the uncertainty.  
Actually, we note the edge of the $A_{\rm CP}^{b\to s\gamma}$ error band 
is drawn in Fig.~\ref{fig:LLRR}.  
In addition, we take $\Delta M_s = 50$ ps$^{-1}$ with 10 \% error as
explained above.  

The squark mixings are determined by $S_{\phi K_S}$, $S_{\eta' K_S}$ 
and $S_{K^*\gamma}$.  As noted above since their dependences to the 
squark mixings are different, the MIA parameters can be measured.  
There is a region where the uncertainty band of all these observables overlaps 
with each other, which is filled by yellow and enclosed 
by red solid lines in the figures.  From Fig.~\ref{fig:LLRR} we find that 
for the luminosity 5 ab$^{-1}$ the squark mixings are determined with an 
error of $20 - 40$ \%, on the other hand, the result is improved when 
the luminosity is 50 ab$^{-1}$ and its uncertainty becomes about 10 \%.  

FCNC is also induced by the LR and RL mixings.  We consider these 
mixings as an origin of flavor mixings instead of the LL and RR case.  
Since we focus on the dipole-penguin dominant processes, as stressed 
above the contours of these processes are found to be nearly the same as 
the LL and RR ones except for the scale of the axes, 
which are approximately rescaled by $c$'s in Eq.~(\ref{eq:totalmixing}).  
Actually the contour lines in Fig.~\ref{fig:LLRR} just shift about 
30 \% at most for the rescaled axes.  As a result the error estimation 
of the resultant squark mixings becomes the same as that of LL and RR, 
and we obtain the uncertainties of $20 - 40$ \% for 5 ab$^{-1}$ and 
about 10 \% for 50 ab$^{-1}$.  

The mixture case of the LL, RR, LR and RL squark mixings is more complex.  
When the contributions from these four mixings are comparable, 
the uncertainty width is larger than the result estimated above.  
This is because the coefficients $c$'s do not satisfy with 
$c^{8G}/c^{7\gamma} ( = \tilde c^{8G}/\tilde c^{7\gamma} ) = 1$ exactly.
Therefore, if we have no information on the ratios of 
the squark mixings, the error bands become about 30 \% wider.  
In that case, the total mixings are consequently determined with the 
uncertainties about 40 \% at the luminosity of 50 ab$^{-1}$.  

\begin{figure}[ht]
  \begin{center}
    \includegraphics{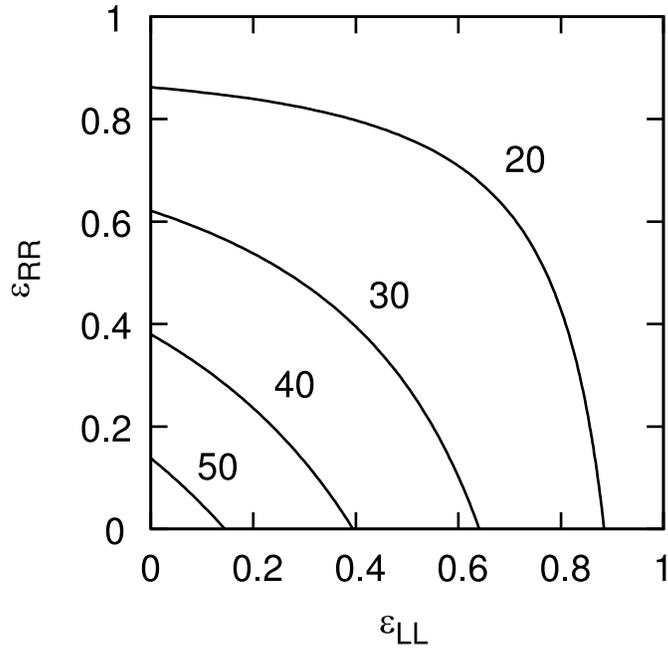}
  \end{center}
  \caption{The constant contours of the $B_s-\overline{B_s}$ mixing, 
    $\Delta M_s$.  The axes are $\epsilon_{LL} = c^{8G} (\delta^d_{LR})_{23}/
    (\delta^d_{LL})_{23}^{\rm (tot)}$ and $L \leftrightarrow R$ for 
    $\epsilon_{RR}$.  Here $(\delta^d_{LL,RR})_{23}^{\rm (tot)}$ of 
    the chromo-magnetic penguin are fixed.}
  \label{fig:DeltaMs}
\end{figure}

In order to identify the four squark mixings, the $B_s-\overline{B_s}$ 
mixing is useful because $\Delta M_s$ is controlled by the product of 
the LL and RR squark mixings.  Figure~\ref{fig:DeltaMs} shows the constant 
contours of $\Delta M_s$ for varying the ratio of LR (RL) component.  
On the other hand, in this figure the total mixings 
$(\delta^d_{LL,RR})_{23}^{\rm (tot)}$ are fixed and as a result 
the other observables in the Fig.~\ref{fig:LLRR} change little.  
Hence from $\Delta M_s$ we obtain one relation for the squark mixings 
in the fixed total mixings.  

In particular, we can definitely distinguish the LL+RR case from the 
others by observing large $\Delta M_s$.  Furthermore when both the 
LL and RR mixings are dominant, $\Delta M_s$ is helpful to measure 
those mixings more precisely.  In fact from Fig.~\ref{fig:LLRR} 
the error band is very narrow even when we consider the theoretical 
uncertainty.  This is striking when we measure the squark mixings 
at the lower luminosity of 5 ab$^{-1}$.  
On the other hand, the mixture case is more complex and we need 
detailed analysis.  

In the above discussion we assumed the condition that the phase of all squark 
mixings align to each other.  One of the reasons comes from the experimental 
constraint from the EDMs.  Although the LL (LR) squark mixing is forced 
to align with RR (RL), when both the RR and RL mixings are very suppressed 
the relative phase between LL and LR mixings becomes loose.  
And the exchanged case of $L \leftrightarrow R$ is also true.  
Then the phase may induce larger uncertainty in the squark mixings.  
However even in such a special case, since we are interested in the situation 
that anomaly of $S_{\phi K_S}$ is large here, the phase is favored to be 
maximal.  Consequently, we reproduce the above result.  

A different choice of center values of the observables modifies the result.  
The error band of $S_{K^*\gamma}$ becomes wide in contrast to $S_{\phi K_S}$ 
and $S_{\eta' K_S}$ when $(\delta^d_{RR})_{23}^{\rm (tot)}$ increases.  
Furthermore we will not expect sizable $A_{\rm CP}^{b\to s\gamma}$ there.  
As a result, in such a region the total RR squark mixing may be measured 
with larger uncertainty, though the total LL squark mixing is again 
determined at the same level in Fig.~\ref{fig:LLRR}.  

We comment on the dependence of the result to the other model 
parameters and the effects from the real part of the squark mixings.  
First let us consider other sets of the soft masses.  
When the gluino, the scalar bottom and the scalar strange become heavy, 
all the $b \to s$ transition amplitudes are suppressed and vice versa.  
As a result, the uncertainties of the squark mixings change little.  
On the other hand, the chargino contributions to the hadronic EDMs have 
different dependence to the soft masses.  They are enhanced by smaller up-type 
squark and chargino masses, contrary to the other observables and constraint.  
In such a case, the LL squark mixing is bounded strongly and the uncertainty 
becomes large.  Also the higgsino mass parameter enhances the gluino 
contributions to the observables, but do not the chargino contributions 
too much.  Therefore the region of smaller $\mu_H$ may again suffer from 
the hadronic EDMs.  We note that the LR mixing is however free 
from the constraint.  Second, other choices of the masses of the chargino, 
the neutralino and the charged Higgs modify the real part 
of the Wilson coefficients.  However the effect is weak and 
we obtain the almost same result.  Finally, the future measurement of 
$S_{\phi K_S}$ may inform us that the squark mixings are not restricted 
to be pure imaginary.  Then the non-vanishing real part generally 
makes the uncertainties large.  In such a case, 
we need more information on the model parameters by some additional 
observables.  

Finally we touch on the flavor-changing neutral Higgs boson decays into the 
bottom and strange quarks.  In the SM, the $W$ boson exchange diagram 
contributes 
to the mode, and the result is found to be very small.  On the other hand, 
the gluino exchange diagram with the squark mixing generally induces a 
flavor-changing coupling of the Higgs.  The authors in Ref.~\cite{Bejar:2004rz} 
showed that the supersymmetry contribution can overshoot that of the SM.  
In fact it becomes larger by about three orders of magnitude when the squark 
mixing takes the value used in the above analysis.  Then such a large 
branching ratio of the flavor-changing Higgs decay may be detected at LHC.  


When LHC starts running and confirms supersymmetry, we come to the next 
stage, that is, to determine the parameters in supersymmetry models and 
to investigate physics at higher energy.  
Super $B$ factory is very appropriate to measure the squark mixings, 
which induce FCNC and CP violation.  
In this letter we studied potential to constrain and measure the squark 
mixings at super $B$ factory.  With the prospected data after one year of 
running we found the mixings is determined at about 10 \% level at best.  
However this result is based on 
theoretical improvements of estimation and on some assumptions.  
Thus we need more rigorous studies.  

\section*{Acknowledgment}
M.E. thanks the Japan Society for the Promotion of Science for financial 
support.
The work of S.M. was supported in part by the Grants-in-aid from the Ministry
of Education, Culture, Sports, Science and Technology, Japan, No.14046201.

\clearpage

\clearpage
\begin{table}[htb]
  \begin{center}
    \begin{tabular}{rcc}
      \hline
      Observable & 5 ab$^{-1}$ & 50 ab$^{-1}$ \\
      \hline
      $A_{\rm CP}^{b\to s\gamma}$ & $0.011$ & $0.005$ \\
      $S_{K^{*0}\gamma}$ & $0.14$ & $0.04$ \\
      $\Delta S_{\phi K_S}$ & $0.079$ & $0.031$ \\
      $\Delta S_{\eta' K_S}$ & $0.049$ & $0.024$ \\
      \hline
    \end{tabular}
  \end{center}
  \caption{Experimental errors after one year of running at super 
    KEKB~\cite{Akeroyd:2004mj}.}
  \label{tab:superKEKB}
\end{table}


\end{document}